\documentstyle[aps,pra]{revtex}

\begin{document}

\twocolumn
\wideabs{

\title{Quantifying entanglement with probabilities}
\author{Markus A. Cirone}
\address{Abteilung f\"ur Quantenphysik, Universit\"at
Ulm, 89069 Ulm, Germany}
\date{\today}
\maketitle

\begin{abstract}
We propose a new approach to the problem of defining
the degree of entanglement
between two particles in a pure state with Hilbert
spaces of arbitrary finite dimensions. The central idea
is that entanglement gives rise to correlations between
the particles that do not occur in separable states.
We individuate the contributions of these correlations
to the joint and the conditional probabilities of
local measurement outcomes. We use these probabilities
to define the measure of entanglement. Our measure
turns out to be proportional to the so--called
2--entropy and therefore satisfies the property
required for any measure of entanglement.
We conclude with an outlook on
the problem of extending our approach to the case of
multipartite systems and mixed states.
\end{abstract}

\vspace{0.5cm}

PACS number(s): 03.65.Ud, 03.65.-w, 03.67.-a
}

\section{Introduction}   

Entanglement is the most surprising
nonclassical property of composite quantum systems
\cite{schr}.
In 1935 Einstein, Podolsky and Rosen (EPR)
pointed out a conflict between their concepts of realism and
completeness of quantum mechanics when two particles
are in an entangled state \cite{epr}. Several years later
Bell revisited the EPR problem considering the case
of particles with Hilbert spaces of finite dimensions
and found his celebrated inequalities \cite{bell}
that show that entangled particles can possess
correlations that cannot be explained within a
local realistic theory.
Such correlations display the essential quantum nature
of the composite system. Many years later Greenberger,
Horne and Zeilinger \cite{ghz} extended Bell's
arguments to three--particle systems.

In the last decade there has been a revival of interest
in entanglement as a consequence of the birth of quantum
information theory \cite{book,nich,bez}. It has been
shown that entangled pairs are a more powerful resource
than separable, i.e., non entangled, pairs in a number
of applications, such as quantum cryptography \cite{eker},
dense coding \cite{dens}, teleportation \cite{tele}
and investigations of quantum channels \cite{us,palm}.

The superior potentiality of entangled states has raised
the question ``How much are two (or more) particles
entangled?'', since pairs with a high degree of
entanglement should be a better resource
than less entangled ones.
Many proposals for a measure of
entanglement have appeared in the last years \cite{enta} but,
in spite of our improved understanding of the problem,
there is no universally accepted measure for the most
general case of $N$ particles in a mixed
state and it is even argued by some authors
that a unique definition does not exist.

The case of two particles $A$ and $B$ in a pure state
$\mid \psi_{\rm AB} \rangle$
has been thoroughly investigated.
Among the measures of entanglement that have
been proposed for bipartite pure states
the entropy of entanglement

\begin{equation}
S(\psi_{\rm AB})\equiv -{\rm Tr_{A}}\left[ \rho_{\rm A}
\ln \rho_{\rm A} \right],
\label{entr}
\end{equation}
where $\rho_{\rm A}\equiv {\rm Tr_{B}}\left(\mid \psi_{\rm AB}\rangle
\langle \psi_{\rm AB} \mid \right)$, is the most popular since it
has a clear physical meaning: It is equal to the ratio $m/n$ of the
number $m$ of maximally entangled states that can be extracted from
$n>m$ identical pure states, in the asymptotic limit
$n\rightarrow \infty$ \cite{benn}.
In fact, some authors claim that the entropy
of entanglement is the only measure of entanglement for
pure states \cite{pope} and that the measure for mixed
states should contain it.
However, there is no universal agreement on this point
\cite{vepl,vida,zycz}. Moreover, the entropy of entanglement
fails to be a measure of entanglement for mixed states \cite{woot}.

In pursuing a measure of entanglement for mixed states and/or
multipartite systems one can follow different approaches.
One can look for measures that contain the entropy of
entanglement Eq. (\ref{entr}) as a special case or examine the
simplest case of pure states from a different point of view in
order to get some hint for the solution of the general problem.
In the latter approach, that we follow here, one can obtain measures
different from the entropy of entanglement, though equivalent
to it.

In the present paper we present a new approach, similar in spirit
to Bell's inequalities, to define the degree of entanglement
of two particles A and B, possessed by
two
observers Alice and Bob, respectively, that are in a pure state
$\mid \psi_{\rm AB} \rangle$. The
Hilbert spaces ${\cal H}_{\rm A}$ and ${\cal H}_{\rm B}$
of the two particles have arbitrary and finite dimensions.
Since several measures of entanglement for
pure states have been already presented by different authors,
it seems opportune to motivate our
proposal. For this purpose, we quote some
remarks about the entangled state

\begin{equation}
\mid \psi_{\rm AB}^{(+)} \rangle =\frac{1}{\sqrt{2}}
\left( \mid \uparrow_{\rm A} \downarrow_{\rm B} \rangle
+ \mid \downarrow_{\rm A} \uparrow_{\rm B} \rangle \right)
\end{equation}
for two spin--1/2 particles that appear in the literature
as an introduction into the basic feature of
entanglement (emphases are ours): ``If Alice performs a
{\em measurement} of spin \ldots she can {\em predict}
with certainty what Bob will {\em measure}'' \cite{nich},
``\ldots neither of the two qubits carries a definite
{\em value}, but \ldots as soon as one of the qubits
is subject to a {\em measurement}, \ldots the other one
will be immediately found to carry the opposite {\em value}''
\cite{bez}. Einstein, Podolsky and Rosen
also say ``If \ldots we can {\em predict} with certainty
the {\em value} of a physical {\em quantity} \ldots'' \cite{epr},
although they comment a different entangled state.

All these remarks concern ({\em a}) measurements of physical
quantities, ({\em b}) results of conditional or joint
measurements and ({\em c}) properties of the state
$\mid \psi_{\rm AB}^{(+)}\rangle$, from which we obtain
the statistics of measurements. In order to define
a measure of entanglement, it is highly
desirable to express these qualitative considerations
touching the heart of entanglement in a quantitative
form. This can be done when one considers that the properties
of of the state $\mid \psi_{\rm AB}^{(+)} \rangle$ discussed above
are a consequence of the correlations that are (exclusively)
due to entanglement. These correlations appear in the
conditional and joint probabilities of measurement outcomes,
as we shall show.

The outline of the paper is as follows:
In Sec. II we define entangled and
separable states and review the properties that a measure
of entanglement must fulfill. We also remind two
mathematical properties, the Schmidt decomposition and
the definition of majorization, and a remarkable connection
between them that has been recently found by Nielsen \cite{niel}.
In Sec. III we examine two two--level
particles. We show that entanglement has consequences
on the conditional probabilities or, equivalently, on the
joint probabilities that given values are obtained
in local measurements on the two particles. The properties of
these probabilities in entangled pairs result
radically different than in separable pairs. We use
this difference to define the measure of entanglement.
We extend then our measure to $N$--level systems.
Our measure results equal, up to a normalizing factor,
to the 2--entropy \cite{horo,krau},
which is related to the 2--R\`enyi entropy \cite{reny}.
Our approach can be extended without changes to two particles
with different numbers $N$ and $N'$ of levels.
Our findings can thus be considered as a physical motivation
in favour of the 2--entropy as a measure of entanglement.
In Sec. IV, after summarizing our findings, we comment the
problems of defining the degree of entanglement in
multipartite systems and in mixed states and generalizing
our approach to these cases.

\section{Prolegomena}

Two particles in a pure state are said to be separable
if their state $\mid \psi_{\rm AB} \rangle$ can be
decomposed into the tensor product

\begin{equation}
\mid \psi_{\rm AB} \rangle = \mid \phi_{\rm A} \rangle
\otimes \mid \phi_{\rm B} \rangle
\end{equation}
of two one--particle states $\mid \phi_{\rm A} \rangle$
and $\mid \phi_{\rm B} \rangle$. In the more general case
of mixed states described by a density matrix
$\rho_{\rm AB}$ the particles are said to be separable
if one can write

\begin{equation}
\rho_{\rm AB} = \sum_{i} p_{i} \rho^{i}_{\rm A}
\otimes \rho^{i}_{\rm B},
\end{equation}
where $\rho^{i}_{\rm A}$ ($\rho^{i}_{\rm A}$) are density
matrices for particle A (B), $\sum_{i}p_{i}=1$ and
$p_{i}\ge 0$ for any $i$. If the particles are not
separable, they are said to be entangled.

What conditions must the measure of entanglement
$E(\rho)$ ($E(\psi)$ for pure states)
satisfy? Clearly we want ({\em i}) $E(\rho)=0$ if
and only if
the particles are separable. Moreover, $E(\rho)$
must be positive, so with the introduction of
appropriate normalization factors one can also
require $0 \le E(\rho) \le 1$ for
any state $\rho$, though this is not compulsory.

The second requirement is that ({\em ii}) $E(\rho)$ must
be invariant under the action of
local unitary operations $U_{\rm A}$ and $U_{\rm B}$,
i.e., $E(\rho)=E(U_{\rm A}U_{\rm B} \rho
U^{\dagger}_{\rm A}U^{\dagger}_{\rm B})$. In other words,
the amount of entanglement cannot depend on the choice
of basis.

A more subtle property comes from the
observation that local general measurements (LGM), i.e., measurements
performed by Alice and Bob on each separated
particle cannot increase on average the amount of entanglement
between A and B, not even when the measurements
on A and B are correlated after exchange of classical
information between Alice and Bob \cite{vepl}.
Indeed, entanglement is a nonlocal property
of quantum systems and therefore it can only deteriorate
or remain constant when the system is locally probed.
Local general measurements
are described by sets of operators
$\{ A_{i} \}$,
 $\{ B_{i} \}$, acting on particles A and B,
respectively,
that satisfy the completeness relations
$\sum_{i}A_{i}^{\dagger}A_{i}=1$
and $\sum_{i}B_{i}^{\dagger}B_{i}=1$. After an LGM
assisted by classical communication (CC) has been performed,
the pair goes into the state

\begin{equation}
\rho_{k}=A_{k}B_{k}\rho A_{k}^{\dagger} B_{k}^{\dagger}
\end{equation}
with probability $p_{k}$. Therefore
({\em iii}) entanglement monotonicity \cite{vida}

\begin{equation}
E(\rho) \ge \sum_{k}p_{k}E(\rho_{k})
\label{mono}
\end{equation}
for the degree of entanglement $E$ is required, which ensures
that on average LGM+CC do not increase entanglement.

The three conditions ({\em i})--({\em iii}) presented
in \cite{knig,vepl} are the most frequent conditions
reported in the literature. It has also been argued
that condition ({\em iii}) of monotonicity
alone entails conditions ({\em i}) and ({\em ii})
\cite{vida}. Nonetheless, we keep the conditions
({\em i}) and ({\em ii}) since their meaning is easy
to understand and they are often easy to verify, so that
they constitute a first test for any definition of $E$.

We recall now three mathematical properties concerning
the pure states $\mid \psi_{\rm AB} \rangle$.
The first is the Schmidt decomposition
\cite{knek}. Assuming ${\rm dim} {\cal H}_{\rm A}=N
\le N'={\rm dim} {\cal H}_{\rm B}$ it is always
possible to define a basis $\{ \mid i_{\rm A} \rangle \}$,
$\{\mid i_{\rm B} \rangle \}$ in
each Hilbert space such that

\begin{equation}
\mid \psi_{\rm AB} \rangle = \sum_{i=1}^{N} \sqrt{\lambda_{i}}
\mid i_{\rm A} \rangle \otimes \mid i_{\rm B} \rangle,
\end{equation}
where the Schmidt parameters $\lambda_{i}$ are the
eigenvalues of the single--particle reduced density
matrices $\rho_{\rm A}\equiv {\rm Tr}_{\rm B}(\rho_{\rm AB})$
or $\rho_{\rm B}\equiv {\rm Tr}_{\rm A}(\rho_{\rm AB})$
obtained after tracing over the other particle. Since
the Schmidt parameters $\lambda_{i} $ are invariant
under a change of basis, they are good ingredients
for any measure of entanglement \cite{woot}.

The second property comes from majorization
theory \cite{bhat}. Given two real $N$--dimensional
normalized vectors $x \equiv (x_{1},\ldots , x_{N})$
and $y \equiv (y_{1},\ldots , y_{N})$ with their
components in decreasing order $x_{1} \ge \ldots \ge x_{N}$,
$y_{1} \ge \ldots \ge y_{N}$, the vector $x$ is said to be
majorized by $y$, written $x \prec y$, if

\begin{equation}
\sum_{i=1}^{k} x_{i} \le \sum_{i=1}^{k} y_{i}
\end{equation}
for any $k=1, \ldots , N$. The functions that preserve
majorization are called Schur convex \cite{zycz} whereas
those that reverse majorization are said to be Schur
concave. In particular, the function
$f_{q}(x)=\sum_{i=1}^{N}x^{q}_{i}$ is Schur convex for any
$q\ge 1$ and therefore $1-\sum_{i=1}^{N}x^{q}_{i}$ is
Schur concave.

The third and last property we recall has been recently
demonstrated by Nielsen \cite{niel}. If the pure state
$\mid \psi_{\rm AB} \rangle$ is transformed into
another pure state $\mid \phi_{\rm AB} \rangle$ by
means of LGM + CC, the Schmidt parameters
$\lambda(\phi) \equiv (\lambda_{1}(\phi), \ldots,
\lambda_{N}(\phi))$ of the state $\mid \phi_{\rm AB} \rangle$
majorize the analogous parameters
$\lambda(\psi) \equiv (\lambda_{1}(\psi), \ldots,
\lambda_{N}(\psi))$ of the initial state
$\mid \psi_{\rm AB} \rangle$.
As a consequence, all
entanglement measures expressed in terms of the
Schmidt parameters that are also Schur concave satisfy
the entanglement monotonicity condition Eq. (\ref{mono}).

\section{Probabilities and measure of entanglement}

We first discuss the case of two two--level systems
in a pure state with great detail.
We move then to the more general
case of two $N$--level systems in a pure state.
Since our starting point is the Schmidt decomposition
our results can be immediately extended to the case when
the Hilbert spaces of the two particles have different
dimensions.

\subsection{$2 \times 2$ systems}    

When the particles A and B have only two levels
$\mid 1 \rangle$ and $\mid 2 \rangle$ the Schmidt
decomposition for any pure state
$\mid \psi^{(2)}_{AB}\rangle$ reads

\begin{equation}
\mid \psi^{(2)}_{AB}\rangle = \sqrt{\lambda}
\mid 1_{\rm A} 1_{\rm B} \rangle +\sqrt{1-\lambda}
\mid 2_{\rm A} 2_{\rm B} \rangle,
\label{psi2en}
\end{equation}
where $0\le \lambda \le 1$ is the only parameter.
Although the Schmidt decomposition can be not
uniquely defined when $\lambda=1/2$ \cite{knek},
the Schmidt coefficients $\sqrt{\lambda}$ and
$\sqrt{1-\lambda}$ are unique for a given state
$\mid \psi_{\rm AB} \rangle$.

When Alice (Bob) performs a measurement that projects
$\mid \psi^{(2)}_{AB}\rangle$ on the basis states,
she (he) finds her (his) particle in state
$\mid 1 \rangle$ with probability

\begin{equation}
P(1_{\rm A})=P(1_{\rm B})=\lambda
\label{p1}
\end{equation}
and in state $\mid 2 \rangle$ with probability

\begin{equation}
P(2_{\rm A})=P(2_{\rm B})=1-\lambda.
\label{p2}
\end{equation}
From these measurements alone Alice and Bob cannot conclude
whether their particles are entangled or not,
since they would obtain the same results
Eq. (\ref{p1}) and Eq. (\ref{p2}) for simple measurement
probabilities with the separable pair

\begin{eqnarray}
\mid \psi^{(2)}_{AB}\rangle_{\rm sep} & \equiv & \left(
\sqrt{\lambda} \mid 1_{\rm A} \rangle +\sqrt{1-\lambda}
\mid 2_{\rm A} \rangle \right) \nonumber \\
& \otimes & \left( \sqrt{\lambda}
\mid 1_{\rm B} \rangle +\sqrt{1-\lambda}
\mid 2_{\rm B} \rangle \right).
\label{psi2di}
\end{eqnarray}

However, they can obtain hints about entanglement
when they exchange information about their measurements.

Entanglement manifests itself in the conditional probabilities
$P(n_{\rm A} \mid m_{\rm B})$ of finding particle A in the
state $\mid n \rangle$ after that particle B has been
found in the state $\mid m \rangle$, and in the
joint probabilities $P(n_{\rm A},m_{\rm B})$ of finding
particle A in state $\mid n \rangle$ and particle B in
state $\mid m \rangle$. Indeed, from the state Eq. (\ref{psi2en})
we find

\begin{eqnarray}
P(1_{\rm A} \mid 1_{\rm B})= & 1 & =P(2_{\rm A} \mid 2_{\rm B}),
\nonumber \\
P(1_{\rm A} \mid 2_{\rm B})= & 0 & =P(2_{\rm A} \mid 1_{\rm B}).
\end{eqnarray}
For the separable state Eq. (\ref{psi2di}) it results
$P(n_{\rm A} \mid m_{\rm B})=P(n_{\rm A})$, since measurements
on particle B have no influence on the state of particle A
and cannot change the statistics of measurements
performed on A. Therefore, the inequality 
$P(n_{\rm A} \mid m_{\rm B})\neq P(n_{\rm A})$ is a consequence
of entanglement. Analogous considerations can be done
for the joint probabilities $P(n_{\rm A}, m_{\rm B})$.
For the separable state Eq. (\ref{psi2di}) the joint
probability factorizes into the product of simple
probabilities $P(n_{\rm A},m_{\rm B})=P(n_{\rm A})P(m_{\rm B})$,
since the results of local measurements on separable
pairs are independent events, whereas for the entangled
state Eq. (\ref{psi2en}) we find

\begin{eqnarray*}
P(1_{\rm A},1_{\rm B})& = & \lambda \neq
P(1_{\rm A})P(1_{\rm B})=\lambda^2, \\
P(2_{\rm A},2_{\rm B}) & = & 1-\lambda \neq
P(2_{\rm A})P(2_{\rm B})=(1-\lambda)^2, \\
P(1_{\rm A},2_{\rm B}) & = & 0 \neq
P(1_{\rm A})P(2_{\rm B})=\lambda(1-\lambda), \\
P(2_{\rm A},1_{\rm B})& = & 0 \neq
P(2_{\rm A})P(1_{\rm B})=\lambda(1-\lambda), \\
\end{eqnarray*}
These considerations on the conditional and joint probabilities
can be merged together and used to quantify the degree
of entanglement when we consider that the difference

\begin{eqnarray}
\mid P(n_{\rm A} \mid m_{\rm B})-P(n_{\rm A}) \mid P(m_{\rm B})
\nonumber \\
=\mid P(n_{\rm A},m_{\rm B})-P(n_{\rm A})P(m_{\rm B}) \mid
\label{diff}
\end{eqnarray}
points out the quantum correlations between measurements
on the two particles. We stress again that these correlations 
are solely due to entanglement. The differences in Eq. (\ref{diff})
can be seen as the differences of conditional or joint
probabilities of the entangled state (\ref{psi2en})
and the separable state
(\ref{psi2di}) that give the same simple probabilities. 

We are now ready to define the measure of the degree of 
entanglement $E$ of state Eq. (\ref{psi2en}) as

\begin{equation}
E(\psi_{\rm AB}^{(2)})\equiv \sum_{n=1}^{2}\sum_{m=1}^{2}
\mid P(n_{\rm A},m_{\rm B})-P(n_{\rm A})P(m_{\rm B}) \mid,
\label{en2}
\end{equation}
where $n$ and $m$ denote states of
the Schmidt basis. Simple calculations
give

\begin{eqnarray}
E(\psi_{\rm AB}^{(2)}) & = & \mid \lambda -\lambda^2 \mid
+ \mid -\lambda(1-\lambda)\mid + \mid -\lambda(1-\lambda)\mid
\nonumber \\
& & +
\mid 1-\lambda-(1-\lambda)^2 \mid
= 4\lambda(1-\lambda) \equiv E(\lambda)
\end{eqnarray}
This expression is the same result given by other two measure
of entanglement, the Bures metric \cite{vepl} and the
2-entropy and therefore
satisfies the conditions ({\em i})--({\em iii}).

\subsection{$ N \times N$ systems}

Our definition Eq. (\ref{en2}) for the measure of entanglement
can be easily extended to particles with more
than two levels. When the particles have $N$ levels
$\mid 1 \rangle$, $\mid 2 \rangle$, \ldots, $\mid N \rangle$,
the Schmidt decomposition of their state
$\mid \psi^{(N)}_{\rm AB} \rangle$ reads

\begin{equation}
\mid \psi^{(N)}_{\rm AB} \rangle = \sum_{i=1}^{N}
\sqrt{\lambda_{i}} \mid i_{\rm A} i_{\rm B} \rangle,
\label{psinen}
\end{equation}
where the Schmidt parameters $\lambda_{i}$ satisfy
the normalization condition $\sum_{i}^{N}\lambda_{i}=1$.

In analogy to Eq. (\ref{en2}) we define the measure of
the degree of entanglement $E$ of the state
$\mid \psi^{(N)}_{\rm AB} \rangle$, Eq. (\ref{psinen})
as

\begin{eqnarray}
E(\psi_{\rm AB}^{(N)}) & \equiv & \frac{N}{2(N-1)}
\nonumber \\
& & \times \sum_{n,m=1}^{N}
\mid P(n_{\rm A},m_{\rm B})-P(n_{\rm A})P(m_{\rm B}) \mid
\label{enn}
\end{eqnarray}
where $n$, and $m$ again denote states
of the Schmidt basis and a normalization factor
$N/[2(N-1)]$ has been introduced in order to ensure
$0\le E \le 1$. This normalization
factor reduces to 1 for $N=1$, so the measure Eq. (\ref{enn})
contains the measure Eq. (\ref{en2}) for two--level
particles as a particular case.

After simple calculations we arrive at

\begin{equation}
E\left( \psi^{(N)}_{\rm AB}\right) =\frac{N}{N-1}\left[
1-\sum_{i-1}^{N} \lambda_{i}^2 \right] \equiv
E(\lambda_{1}, \ldots, \lambda_{N}).
\label{exple}
\end{equation}
This expression is proportional to the 2--entropy \cite{krau}
and is therefore a good measure of entanglement.

Indeed, condition ({\em i}) is clearly satisfied.
Since $\lambda_{i}\le 1$,
$E$ is equal to 0 if and only if $\sum_{i=1}^N\lambda_{i}^2=1$.
Because of the constraint $\sum_{i=1}^N\lambda_{i}=1$,
this occurs if one and only one
Schmidt parameter $\lambda_{i}$ is nonvanishing and equal to 1.
This corresponds to the disentangled state

\begin{equation}
\mid \psi^{(N)}_{\rm AB} \rangle \, = \, \mid i_{\rm A} \rangle
\otimes \mid i_{\rm B} \rangle.
\end{equation}

The highest degree of entanglement $E=1$ occurs when
$\lambda_{1}=\ldots =\lambda_{N}=1/N$. Indeed, the function

\begin{equation}
f(\lambda_{1},\ldots, \lambda_{N})=
E(\lambda_{1},\ldots, \lambda_{N})+
\mu \left( \sum_{i=1}^N\lambda_{i}-1 \right)
\end{equation}
where $\mu$ is a Lagrange multiplier has a unique maximum
at $\tilde{\lambda}_{1}=\ldots=\tilde{\lambda}_{N}=1/N$.

Condition ({\em ii}) is also satisfied since the Schmidt
parameters are invariant under a change of basis \cite{woot}.
Finally, condition
({\em iii}) is also satisfied because the
expression Eq. (\ref{exple}) is Schur concave and thus entanglement
monotone.

\section{Discussion and conclusions}

We have examined the peculiar features of an entangled pure
state of a bipartite system. Entanglement is the origin of
correlations between the two particles of the system and
manifests itself in the properties of the conditional
and joint probabilities of measurement outcomes that are defined
with the help of the Schmidt decomposition. These properties can be
used to define a measure of entanglement, that results to be
essentially equal to the 2--entropy.

An important question is how can our approach be extended
to (a) multipartite systems and
(b) mixed states, where the Schmidt decomposition does
not always exist. At first sight these
seem to be two radically different cases.

Let us first consider the case of a multipartite system
composed of three particles A, B, and C, in a pure state
$\mid \psi_{\rm ABC} \rangle$ of a Hilbert space of finite dimension.
In our opinion, one has to be extremely careful in
formulating the correct question. We believe that the
questions ``What is the degree of entanglement of
$\mid \psi_{\rm ABC}\rangle$?'' and
``How much is A entangled with B+C in
$\mid \psi_{\rm ABC} \rangle$?'' are fundamentally different
and only the latter makes clearly sense.
Indeed, if we divide
the system into the two subsystems A and B+C we can still
use our measure
to define the degree of entanglement of A with B+C.
In
general, the measure $E_{\rm A, B+C}$ of entanglement
between A and B+C is different from the measure
$E_{\rm B,A+C}$ of entanglement between B and A+C. It is
thus not clear if a measure of entanglement of the
whole system can ever be defined.
Only if the three--partite system
has a Schmidt decomposition we have
$E_{\rm B,A+C}=E_{\rm B,A+C}=E_{\rm B,A+C}$ and
we can safely take this quantity as a measure of entanglement
of the whole system.

One can also ask `What is the degree of entanglement
between A and B in $\mid \psi_{\rm ABC} \rangle$?'.
One could be tempted to say that such a question is not
well posed. After all, the lesson of EPR's argument is that
entanglement is a non--local, i.e., non--individual
property of a quantum system and
that none of the components of the system can be neglected,
contrarily to what the last question above does.
However, if one takes the density matrix $\rho_{\rm ABC}$
and traces over the particle C, one ends up with the density
matrix $\rho_{\rm AB}$ of the two particles A and B.
In this way the problem of measuring the entanglement
between two particles of a multipartite system in a pure
state seems to reduce to the problem of measuring the
entanglement between two particles in a mixed state.
Although the two physical systems (two particles of a
three--partite system in a pure state vs. two particles
in a mixed state) are conceptually and physically
different, we cannot distinguish between them
if they are described by the same density matrix and
the third particle C is not accessible to us.

We are thus led to the problem of extending our
approach to mixed states. It seems that there are two
major problems. We have already mentioned that the
Schmidt decomposition
does not exist in general for mixed states \cite{barn}.
The second problem is that in mixed states also classical
correlations occur \cite{wern}.
One must then be able to separate
the contribution of classical correlations to the
joint and conditional probabilities from the quantum
correlations that have their origin in entanglement.
These problems are currently under investigation.

I wish to thank G. Alber, I. Cirac, A. Delgado, D. G. Fischer,
M. Freyberger, B. Kraus, M. Ku\'s, M. Lewenstein,
H. Mack, G. Metikas, G. M. Palma,
F. Persico, A. Sanpera and K. \.Zyczkowski for numerous
discussions, interesting comments and
for drawing my attention to some relevant
articles. This work is supported by the IHP program
of the European Union with the network `QUEST'.

\end{document}